\documentclass[prd,twocolumn,nofootinbib]{revtex4}
\usepackage{latexsym}
\usepackage{amsthm}
\usepackage{amsmath}
\usepackage{graphicx}  
\usepackage{bm}        
\usepackage{amssymb}   
\usepackage{hyperref}
\newcommand{\gsim}{\lower.7ex\hbox{$\;\stackrel{\textstyle>}{\sim}\;$}}
\newcommand{\lsim}{\lower.7ex\hbox{$\;\stackrel{\textstyle<}{\sim}\;$}}

\newcommand{\GeV}{\,\mathrm{GeV}}
\newcommand{\MeV}{\,\mathrm{MeV}}

\newcommand{\be}{\begin{equation}}
\newcommand{\ee}{\end{equation}}
\newcommand{\bea}{\begin{eqnarray}}
\newcommand{\eea}{\end{eqnarray}}

\newcommand{\bef}{\begin{figure}[htbp]\begin{center}}
\newcommand{\eef}{\end{center}\end{figure}}

\begin{document}
\title{The Production and Discovery of True Muonium in Fixed-Target Experiments }

\author{Andrzej Banburski}
\affiliation{Perimeter Institute for Theoretical Physics, Waterloo, ON N2L 2Y5, Canada}
\author{Philip Schuster}
\affiliation{Perimeter Institute for Theoretical Physics, Waterloo, ON N2L 2Y5, Canada}

\date{\today}

\begin{abstract}
Upcoming fixed-target experiments designed to search for new sub-GeV forces 
will also have sensitivity to the never before observed True Muonium atom, a bound state
of $\mu^+\mu^-$. We describe the production and decay characteristics of True Muonium relevant 
to these experiments. 
Importantly, we find that secondary production mechanisms dominate over primary 
production for the long-lived 2S and 2P states, leading to total yields an order of magnitude larger than naive estimates 
previously suggested.
We present yield estimates for True Muonium as a function of energy fraction and decay length, useful 
for guiding future experimental studies. 
Discovery and measurement prospects appear very favorable. 
\end{abstract}

\maketitle
\section{Introduction}

A new generation of fixed-target experiments at Jefferson Laboratory and Mainz 
\cite{Abrahamyan:2011gv,Merkel:2011ze,HPS,Hewett:2012ns}
designed to search for new sub-GeV scale forces will also have unprecedented sensitivity to rare QED processes.
Experiments such as the Heavy Photon Search (HPS) \cite{HPS}
will have the capacity to precisely identify displaced vertices in the $e^+e^-$ final state arising 
downstream of an electron beam scattering off a high-Z target. 
This opens the door to discovering True Muonium (TM), the 
QED bound state of a $\mu^+\mu^-$ pair, and studying it for the first time. 
The decay and spectral characteristics of TM can then be 
used to further test properties of the muon, and used to study bound state 
physics, with non-perturbative analogues in QCD, in a calculable regime \cite{Brodsky:2009gx,Brodsky:2011fc}. 
Studying TM is additionally motivated 
in light of long-standing discrepancies between theory and 
observations of $(g-2)_{\mu}$, as well as more recent
discrepancies in the measured proton charge radius in muonic 
hydrogen \cite{Pohl:2010zza,TuckerSmith:2010ra}. 

In this short note, we compute total yields for TM production
in fixed-target experiments like HPS,
properly including primary production, break-up, and excitation reactions
of relativistic TM.
HPS uses a forward peaked electromagnetic calorimeter 
to trigger on coincident charged particles or energetic photons
produced in $e^-$ -nucleus collisions. Then, using a relatively small silicon tracker, 
charged particles and any associated displaced vertices are identified within $\sim 10$ cm
downstream from the target. This allows HPS to look for the production 
of rare QED states with long lifetime, such as TM. 
Our discussion will focus on the total production of triplet states 
$1^3S_1$, $2^3S_1$, and $2^3P_2$ because these states eventually decay
to $e^+e^-$ and can be detected by a silicon tracker. 
 We do not aim to provide a precise prediction for the total TM production rates, 
but rather include all dominant effects of TM production and decay 
to help guide upcoming efforts to discover TM.
In addition to the well-known primary production of TM, we show that secondary 
production via $1^3S_1$ excitations is the main source of $2^3S_1$ and $2^3P_2$ production.
The $2S$ and $2P$ state are especially long-lived, so our finding
is particularly important for fixed-target experiments that can identify vertices 
of $e^+e^-$ pairs in the range of $\sim 1$ cm to several cm. 
$2^3S_1$ and $2^3P_1$ produced via secondary mechanisms 
should be readily discoverable in an HPS-style experiment. 

In Section ~\ref{sec:yields}, we describe our methods, and summarize our
overall calculation and results for the yield of TM events relevant for upcoming 
fixed-target experiments. For illustration, and for ease of comparison with existing literature,
we present specific yield results for a Lead target. The results for Lead are also comparable 
to the results for other common high-Z targets like Tungsten or Tantalum. 
In Section ~\ref{sec:primary}, we describe primary production calculations, 
followed by secondary mechanisms in Section ~\ref{sec:secondary}.
We end with a short discussion regarding discovery and measurement prospects.

\section{Production and Decay Yields} \label{sec:yields}
 

For an HPS-style setup, the triplet $1^3S_1$, $2^3S_1$, and $2^3P_2$ will 
dominate the TM signal.
The singlet configurations decay to $\gamma\gamma$, which 
is very difficult to separate from QED backgrounds. 
Primary production through single-
and 3- photon reactions \cite{Holvik:1986ty,ArteagaRomero:2000yh} 
dominate the total production of the $1^3S_1$ state. 
$2^3S_1$ and $2^3P_2$ primary production is down by an order or magnitude
compared to $1^3S_1$~\cite{Holvik:1986ty,ArteagaRomero:2000yh}. 
Additionally, the dissociation for $2^3S_1$ and $2^3P_2$
is an order of magnitude larger than $1^3S_1$~\cite{Denisenko:1987gr,Mrowczynski:1987gq}, 
which makes primary production (followed by target escape) yields negligible. 
The leading source of $2^3S_1$ and $2^3P_2$ production is through 
secondary reactions initiated by $1^3S_1$ scattering into $2^3P_2$,
and $2^3P_2$ scattering or decaying into $2^3S_1$. 
In what follows, we will use primary production rates for $1^3S_1$,
$2^3S_1$, and $2^3P_2$ previously calculated~\cite{Holvik:1986ty,ArteagaRomero:2000yh},
along with $1^3S_1\rightarrow 2^3P_2$ and $2^3P_2\rightarrow 2^3S_1$ transition 
and $1S,2S,2P\rightarrow X$ excitation into all possible final states cross sections computed 
using the formalism of~\cite{Denisenko:1987gr,Mrowczynski:1987gq}. 

We compute total TM yields in terms of the incident number of 
beam electrons $N_e$, as a function of distance traveled through the target. 
In practice, differential yields are a non-trivial function of $x=E/E_{beam}$.
A convenient dimensionless unit of target thickness is the dissociation length 
(or more precisely, dissociation into all final states length), $l_{1^3S_1\rightarrow X}$ 
of the $1^3S_1$ state. 
Thus we convert length $l$ into $z=\frac{l}{l_{1^3S_1\rightarrow X}}$.
Let $N_e(z)$ be the number of beam electrons (with energy $E_{beam}$) 
as a function of distance $z$, and $N_{1S,2S,2P}$ (dropping the spin labels) the average number of TM states.

For the $1^3S_1$ state, the yield as a function of distance is controlled by primary production
and dissociation. To a good approximation,
\be
\frac{dN_{1S}}{dz} = N_e \frac{\sigma(e^-\rightarrow 1S)}{\sigma(1S\rightarrow X)} - N_{1S},
\ee
where $\sigma(e^-\rightarrow 1S)$ is the cross-section for an electron to scatter off 
the nuclear target (Lead for illustration in the remainder of this paper) and produce 
the $1^3S_1$ state, and $\sigma(1S\rightarrow X)$ is the cross-section for 
for the $1^3S_1$ state to scatter and dissociate. 
For the $2^3S_1$ and $2^3P_2$ states, secondary production mechanisms involving 
$1^3S_1\rightarrow 2^3P_2$ and $2^3S_1\leftrightarrow 2^3P_2$ reactions are important,
leading to a yield evolution well approximated by, 
\bea
\frac{dN_{2S}}{dz} &=& N_e \frac{\sigma(e^-\rightarrow 2S)}{\sigma(1S\rightarrow X)} - N_{2S} \frac{\sigma(2S\rightarrow X)}{\sigma(1S\rightarrow X)} \nonumber \\
&& + N_{2P}\frac{\sigma(2P\rightarrow 2S)}{\sigma(1S\rightarrow X)} \\
\frac{dN_{2P}}{dz} &=& N_e \frac{\sigma(e^-\rightarrow 2P)}{\sigma(1S\rightarrow X)} - N_{2P} \frac{\sigma(2P\rightarrow X)}{\sigma(1S\rightarrow X)} \nonumber \\ 
&& + N_{1S}\frac{\sigma(1S\rightarrow 2P)}{\sigma(1S\rightarrow X)} + N_{2S}\frac{\sigma(2S\rightarrow 2P)}{\sigma(1S\rightarrow X)}. 
\eea
$\sigma(e^-\rightarrow 2S)$ and $\sigma(e^-\rightarrow 2P)$ are the primary 
production cross sections for an electron to scatter of the nuclear target and produce
a $2^3S_1$ and $2^3P_2$ states, respectively,
while $\sigma(2S\rightarrow X)$ and $\sigma(2P\rightarrow X)$ are the dissociation 
cross sections.
$\sigma(2S\rightarrow 2P)$ is the cross-section for the $2^3S_1$
state to scatter off the nuclear target into $2^3P_2$, while 
$\sigma(2P\rightarrow 2S)$ is the reverse reaction. 
These are just Boltzmann equations, and the yields quickly asymptote 
to constants as $z$ exceeds the dissociation length set by,
$l_{1S\rightarrow X}=A/N_A\rho\sigma(1S\rightarrow X)$.
We will only be interested in very thin targets, so beam spreading 
effects are ignored, and $N_e$ can be treated as constant.
We have only included the states (and related transitions) that are populated 
by an amount larger than $5\%$ of the total TM yield. 
Triplet-singlet transitions are significantly suppressed \cite{Denisenko:1987gr,Mrowczynski:1987gq},
so the above evolution equations refer only to triplet states. 
Finally, we note that we actually solve this system
as a function of the energy fraction $x$, which is obtained by replacing 
all cross-sections and yield factors by their differential form. 
We express final results in terms of relative yields,
$\frac{dY_i}{dx}=\frac{1}{N_e}\frac{dN_i}{dx}$ as a function of $x$
as well as total yields $\int_0^1 dx \frac{dY_i}{dx}$.

\begin{figure}[htbp]
\includegraphics[width=3.2in]{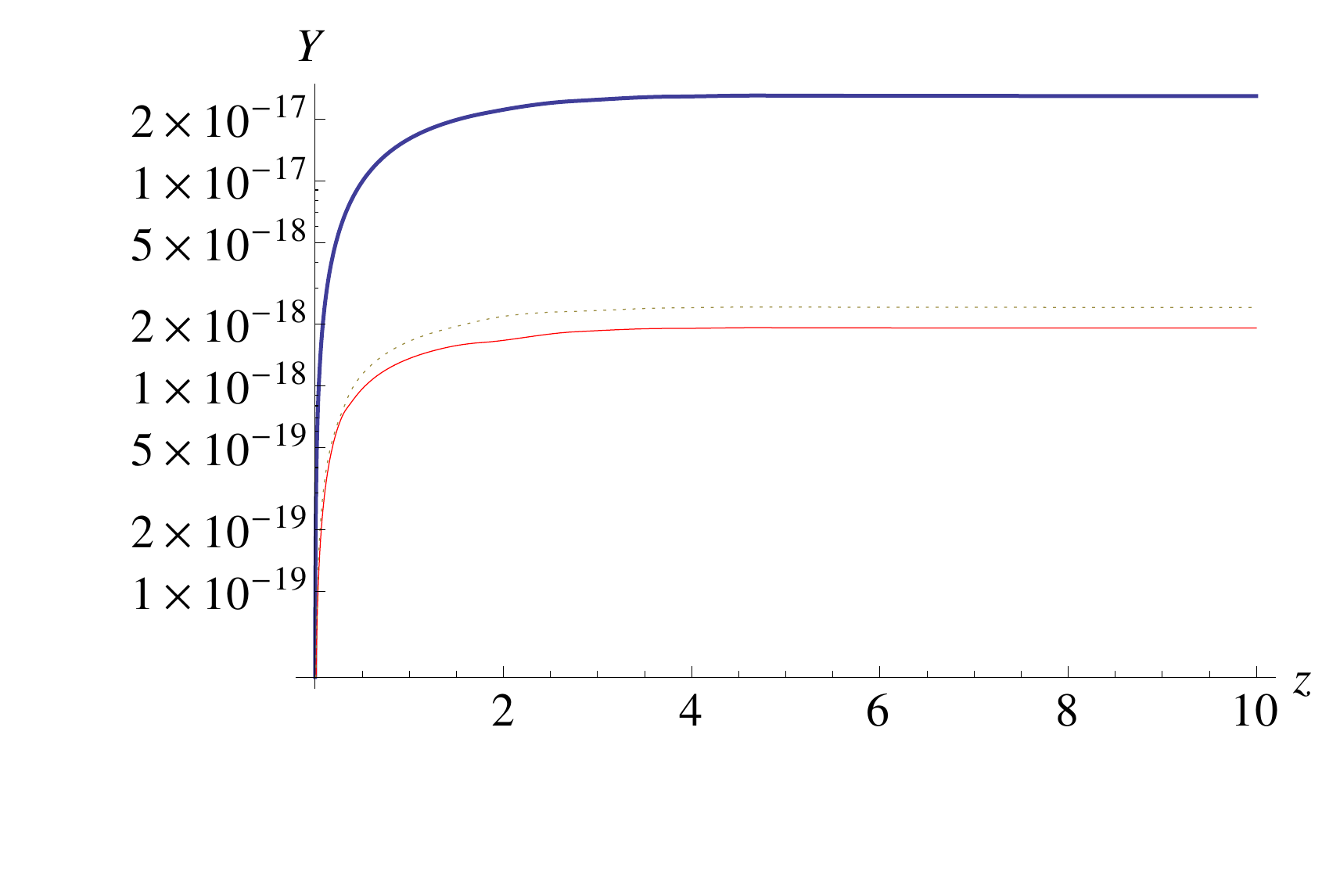}
\caption{\label{Fig:TotalYield} Relative abundance $Y_i=N_i/N_e$ for $i=$$1^3S_1$,$2^3S_1$,$2^3P_2$
TM states as a function of the distance traversed through target,
in units of $z=l/l_{1^3S_1\rightarrow X}$, where $l_{1^3S_1\rightarrow X}$ is the mean free path for $1^3S_1$ breakup.
Blue is $1^3S_1$, red is $2^3S_1$, and dotted gold is $2^3P_2$.}
\end{figure}

Using the results of Sections ~\ref{sec:primary} and ~\ref{sec:secondary}, 
the total relative yields are shown in Figure \ref{Fig:TotalYield} for Lead using a primary beam energy $E_{beam}=6.6 \GeV$. 
For Lead, $l_{1^3S_1\rightarrow X} \approx 4.4\times 10^{-4}$ cm, 
or roughly $10^{-3}$ radiation lengths.
After a few dissociation lengths, the relative yields are constant,
so in practice only the first $10^{-3}$ radiation lengths of the target 
are important for TM production. 
The asymptotic abundance of $1^3S_1$ is controlled
by the total production cross section and the 
dissociation cross section. For $2^3S_1$ and $2^3P_2$,
the asymptotic abundance is controlled by a balance of $1^3S_1\rightarrow 2^3P_2$,
$2^3S_1\rightarrow 2^3P_2$, and $2^3S_1,2^3P_2\rightarrow X$ dissociation reactions. 
In fact, the primary production terms for $2^3S_1$ and $2^3P_2$ are negligible
compared to the secondary production terms, so in practice we can ignore them. 
After a few dissociation lengths, the relative abundances
become approximately constant. For other high-Z materials, the relative abundance
results are very similar -- this is due to the approximate cancellation of the Z-scaling between 
the production cross-section and the dissociation reactions. 



\begin{figure}[htbp]
\includegraphics[width=3.4in]{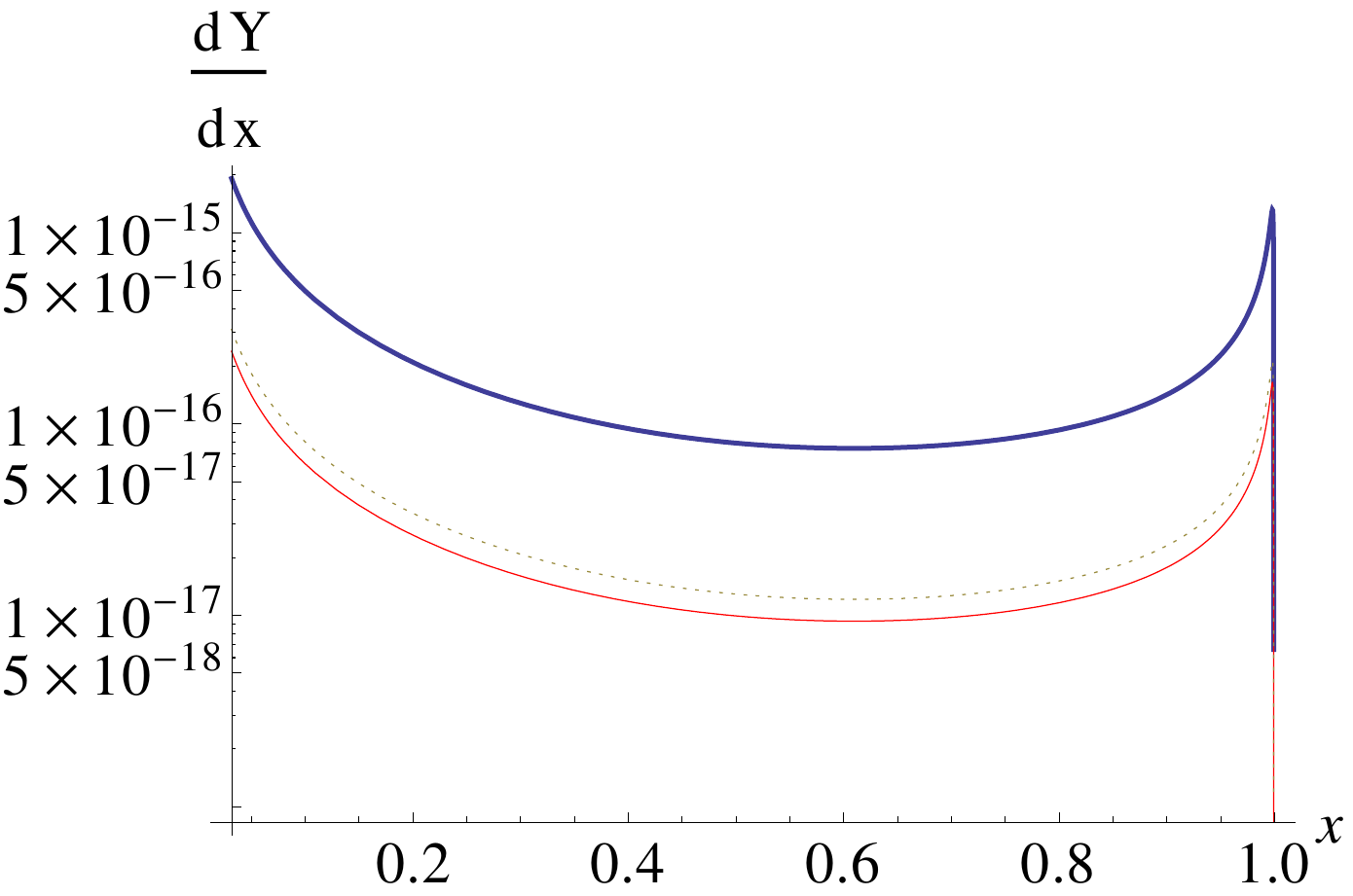}
\caption{\label{Fig:DiffYield} Relative abundance $Y_i=N_i/N_e$ for $i=$$1^3S_1$,$2^3S_1$,$2^3P_2$
TM states at the steady-state configuration, as a function of energy fraction $x=E/E_{beam}$.
$z=10$ was chosen. Blue is $1^3S_1$, red is $2^3S_1$, and dotted gold is $2^3P_2$.}
\end{figure}

In Figure \ref{Fig:DiffYield}, we show
the relative yields as a function of $x$ once 
three dissociation lengths $l_{1^3S_1\rightarrow X}$ of target material have been traversed. 
The dissociation and excitation reactions are approximately 
independent of the TM energy far away from threshold (see Section \ref{sec:primary}). 
Thus, the $x-$dependence is controlled by the primary 
production terms $\frac{d\sigma(e^-\rightarrow 1S,2S,2P)}{dx}$. 
The peak at low-x is dominated by the $3\gamma$ reaction of \cite{ArteagaRomero:2000yh}, 
while the mild high-x peak is dominated by the single $\gamma$
brehmstrahlung-like reaction \cite{Holvik:1986ty}. 

\begin{figure}[htbp]
\includegraphics[width=2.5in]{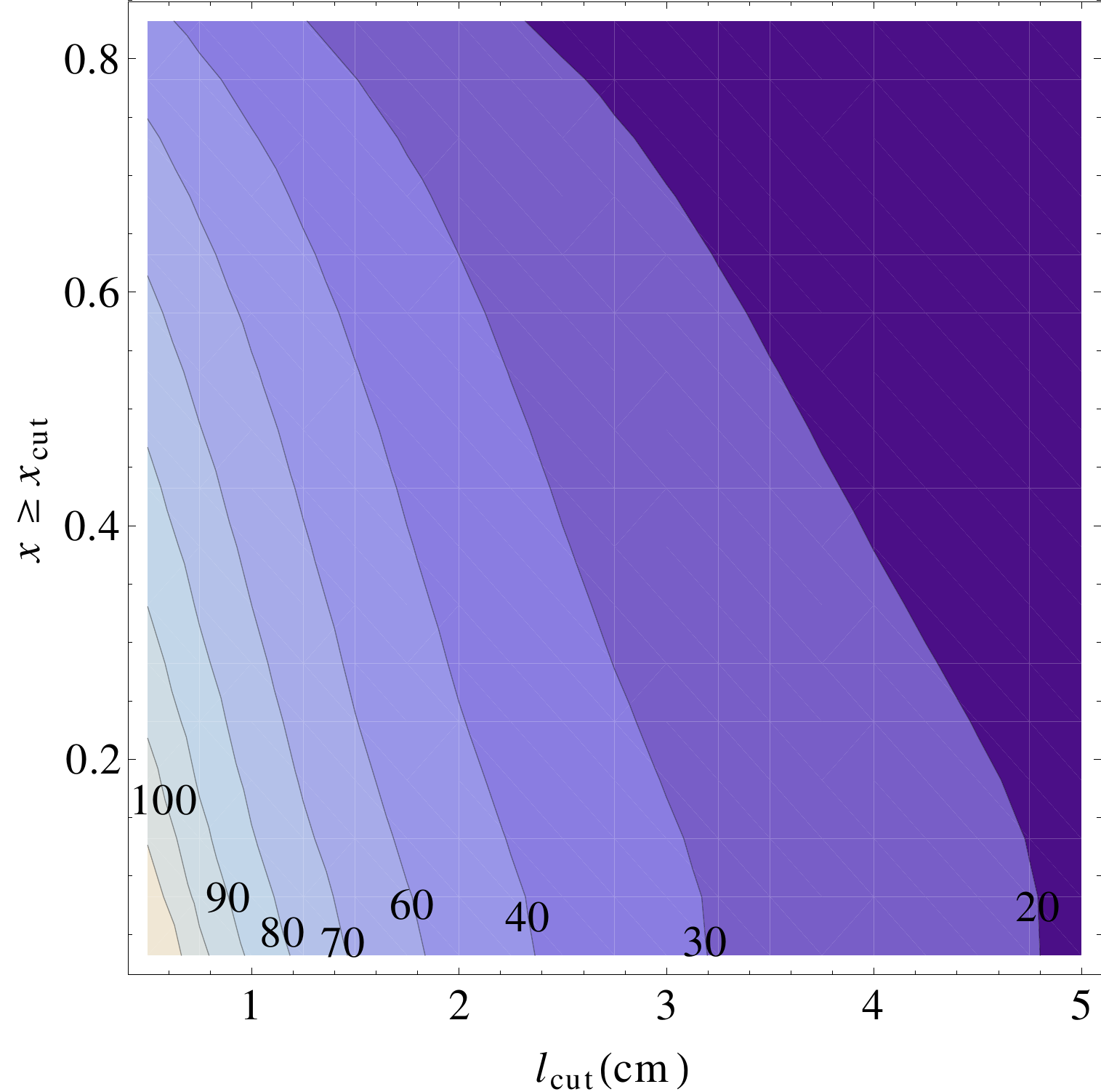}
\caption{\label{Fig:combined} Total yield of $e^+e^-$ events coming from the decays of 
$1^3S_1$,$2^3S_1$, and $2^3P_2$ as a function of energy fraction cut (y-axis) and displaced vertex cut (x-axis in cm).
The yield is given in units of $10^{-18}N_e$, where $N_e$ is the number
of electrons on target, and we've assumed that the target exceeds a thickness of a few $l_{1^3S_1\rightarrow X}$.}
\end{figure}

The primary trigger for HPS-style fixed-target experiments involves 
a cut on the energy fraction of the observed $e^+e^-$ pair, as this is required to 
remove the overwhelming rate of QED $e^+e^-$ background at lower $x=E_{e^+e^-}/E_{beam}$.
An additional cut on the displaced vertex (typically 1-2 cm for $6.6$ GeV beam energies) is
required to remove all background. 
To present yield results in a useful manner, 
we calculate the total yield of TM states as a function
of $x-$cut, $x_c$, and vertex cut, $l_c$, assuming a beam energy of $6.6$ GeV for reference. 
This is just,
\be
Y(x>x_c,l>l_c) = \int_{x_c}^{1} dx \sum_i \frac{dY_i(x)}{dx} \ e^{\left( \frac{-l_c 2m_{\mu}}{xE_{beam} c \tau_i} \right) },
\ee
where $\tau_i$ is the lifetime of the $i^{th}$ TM state. 
In Figure \ref{Fig:combined}, we plot $Y(x>x_c,l>l_c)$ versus 
$x_c$ and $l_c$ for a Lead target and a beam energy of $6.6$ GeV. 
These results are applicable so long as the target is thicker than $\sim l_{1^3S_1\rightarrow X}$. 
We note that the typical opening angle of the $e^+e^-$ pair is $5.8^o \times \left( \frac{6.6 \ GeV}{E_{beam}}  \right)$.

\begin{figure}[htbp]
\includegraphics[width=2.5in]{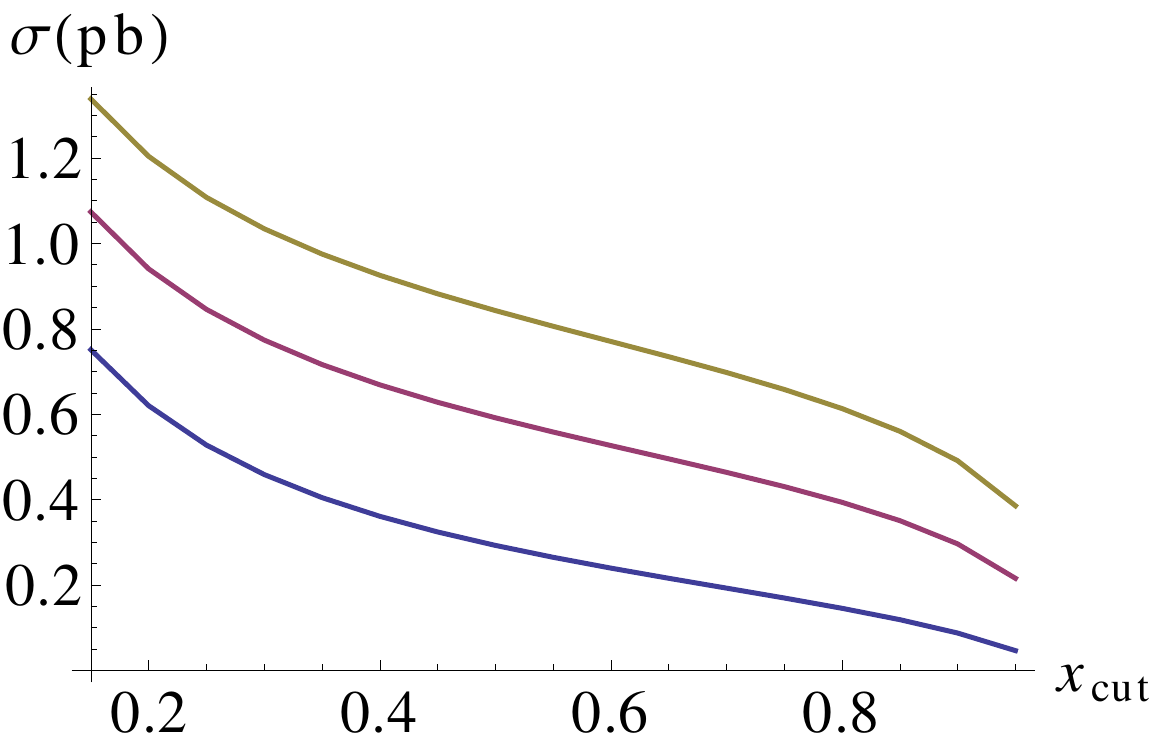}
\caption{\label{Fig:xsecTotal} Total primary production cross-section for all nS states 
as a function of energy fraction cut, $x_{cut}$. The target is Lead, and units are in pb. 
Blue (bottom) is for $E_{beam}=2$ GeV, purple (middle) is for $E_{beam}=6$ GeV, and 
gold (top) is for $E_{beam}=12$ GeV.}
\end{figure}

\begin{figure}[htbp]
\includegraphics[width=2.5in]{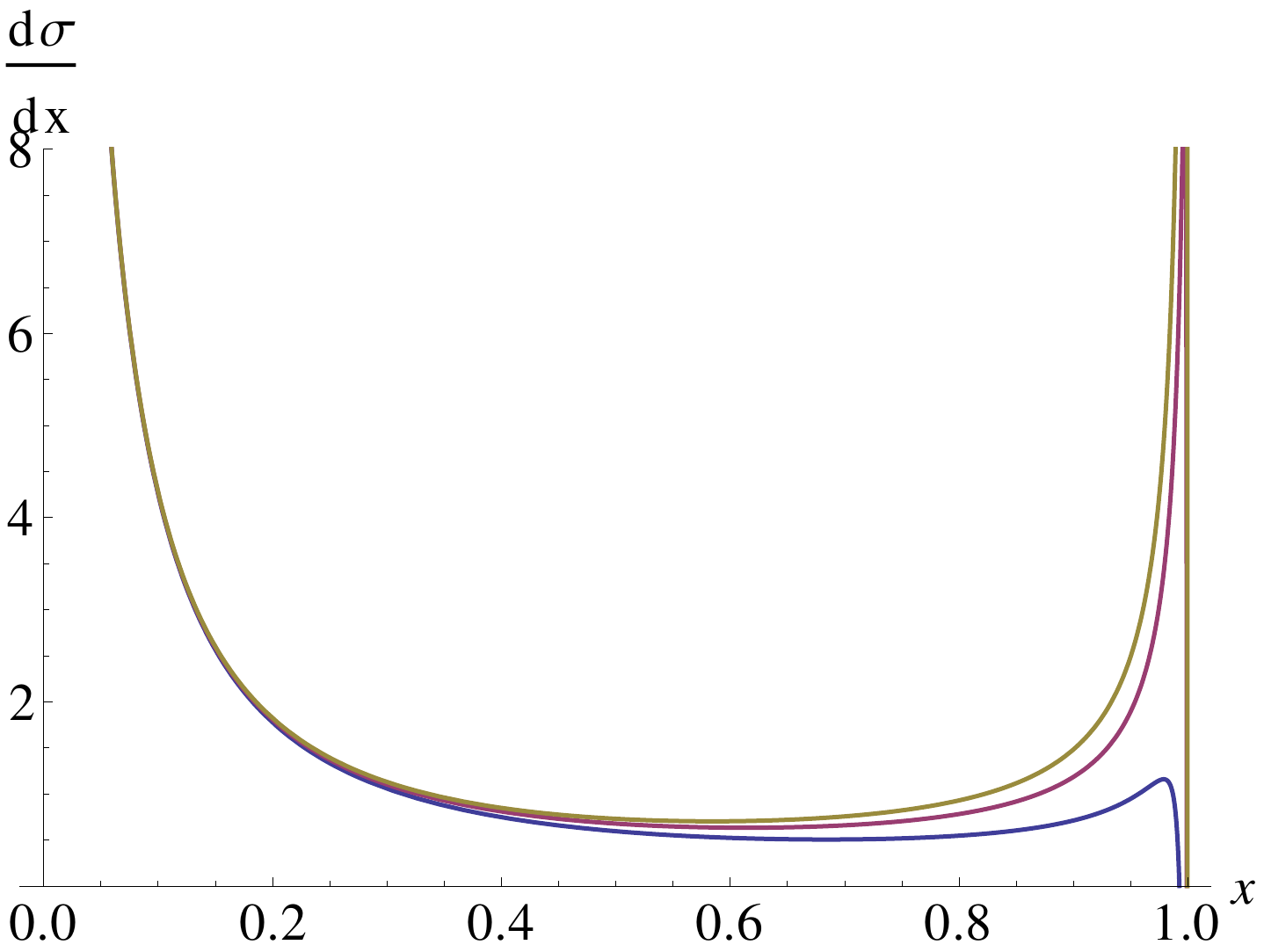}
\caption{\label{Fig:xsecDiff} Differential cross-section for primary production 
of all nS states as a function of energy fraction $x$. The target is Lead, and units are in pb.
Blue (bottom) is for $E_{beam}=2$ GeV, purple (middle) is for $E_{beam}=6$ GeV, and 
gold (top) is for $E_{beam}=12$ GeV.}
\end{figure}

In addition to boosting the decay displacements and altering the typical opening angle, 
the most significant impact of adjusting beam energy is on the primary production cross-section. 
The total TM primary production cross-section as a function of $x_c$ is shown in 
Figure \ref{Fig:xsecTotal}. While the overall growth of the cross-section with energy is mild
(seen by looking at low $x_c$), the cross-section at $x_c\gsim 0.8$ sharply 
increases with energy. For example, the cross-section at $x_c=0.8$ increases 
by a factor of $\sim 3$ going from $E_{beam}=2 \GeV$ to $E_{beam}=12 \GeV$. 
This is mainly due to the energy dependence of the bremstrahlung process,
which both grows and more sharply peaks at high $x$ as energy is increased \cite{Holvik:1986ty,Bjorken:2009mm}.  
Figure \ref{Fig:xsecDiff} illustrates the change in x-dependence with changing beam energy. 

\section{Primary Production}\label{sec:primary}

In this section, we summarize the essential primary production results used in this note. 
The differential production cross section for the Bremsstrahlug process from \cite{Holvik:1986ty} for the n-th energy state is, 
\bea
\text {d$\sigma $} &=& \frac{1}{4 n^3} \frac{Z^2\alpha^7}{m_{\mu}^2}\frac{x\left(1-x\right)\left(1-x+\frac{1}{3}x^2\right)dx}{\left[1-x+(m_e/m_{\mu\bar{\mu}})
^2\right]^2} \times \nonumber \\
&& \ \ \ \ \ \ \ \ \ \left( \ln\left[\frac{\left(E_{beam}/m_{\mu}\right)^2\left(1-x\right)^2}{1-x+(m_e/m_{\mu\bar{\mu}})^2}\right]-1\right),
\eea
where $x = \frac{E_{\mu\mu}}{E_{e}}$, and $E_{beam}$ is the beam energy.
This reaction has a high-x logarithmic divergence regulated by $m_e$, but in practice 
the Weizsacker Williams approximation breaks down near $x\approx 1-m_e/E_{beam}$ \cite{Holvik:1986ty,Bjorken:2009mm}.
Thus, we integrate up to the point where $\frac{\text {d$\sigma $}}{\text{dx}} = 0$ for computing 
the total cross-section. The variation of the cross-section integral that results by varying the cutoff used in this 
procedure is less than $5\%$. 
It is worth noting that the formulas for the total cross-section presented in \cite{Holvik:1986ty} 
are not correct, and in fact run negative for beam energies of $\sim 2 \GeV$. 
This appears to be the result of integrating the Weizsacker Williams form all the way up to $x=1$. 
As a rough check on our results, we compare to a Madgraph calculation for production of a heavy
photon A' \cite{Bjorken:2009mm}, since they both have the same production mechanism. The comparison is based on matching
the kinetic-mixing parameter $\epsilon \equiv g' / e$ (where $g'$ is the coupling of A' to electrons), 
by requiring that the lifetime of a heavy photon of mass 210 MeV
be that of TM. We get a $30\%$ smaller cross-sections compared to our results, but this is within the systematic uncertainties 
of the comparison. 

For the $3 \gamma$ process, we use equation (10) from \cite{ArteagaRomero:2000yh}:
\bea
&&\frac{\text {d$\sigma $}}{\text{dx}} = \text{}\frac{1}{4 n^3} Z^2\frac{\alpha^7}{m_{\mu}^2}\left(\frac{Z\alpha\Lambda}{m_{\mu}}\right)^2\frac{4 B}{x} \times \nonumber \\
&& \left(\left(1- x + \frac{x^2}{2}\right)\text {Log}\left[\frac{(1 - x)\left(m_{\mu}\right)^2}{x^2\left(m_e \right)^2} \right] - 1 + x \right),
\eea
where $B = 0.85$ and $\Lambda = \frac{405}{A^{1/3}} \MeV$. 
We can readily see that there is an additional 
$\frac{Z^2}{A^{2/3}}$ dependence compared to the bremsstrahlung process,
which yields proportionally larger primary production rates in heavier (high Z) targets. 
For the total cross-section, we take the sum of the $3 \gamma$ and bremsstrahlung reactions,
ignoring interference. 
These reactions are peaked in different kinematic regions --- Bremsstrahlung at high-$x$ and $3\gamma$ at low-$x$ ---
so we expect that this approximation is good to $20-30\%$.
In all, we expect that our primary production cross-sections are good to $\sim 40\%$, based on uncertainties 
from atomic form factors, use of the Weizsacker Williams approximation, and interference effects. 

\section{Secondary Production}\label{sec:secondary}

To compute secondary production effects (interaction of TM with matter), we use the formalism of 
\cite{Denisenko:1987gr}, \cite{Mrowczynski:1987gq}, and \cite{Afanasyev:2001zh}. 
In this formalism, TM is treated to be initially in the rest frame, and matter is treated as 
relativistic structureless particles. 
This allows one to tractably consider excitations of TM, from
which we can also calculate break-up cross-section. 
The main contribution to the cross-sections is from electric interactions, while spin-spin, 
para-ortho and magnetic (internal motion of the atom) interactions are smaller by a few orders 
of magnitude, and thus can be neglected.
The more complete calculation of many of these effects 
is presented in \cite{Afanasyev:2001zh}, but for atoms like Pb 
neglecting these leads to $\sim 1\%$ error.
While the $1S$ transition results are included in earlier literature (see \cite{Mrowczynski:1987gq,Denisenko:1987gr}),
the relevant $2S$ and $2P$ transitions are included here for the first time.  

Before we calculate cross-sections, one has to calculate the transfer matrix elements:
\bea
F^{nlm,n'l'm'}(q) &=& \int _0^{\infty }\int _0^{\pi }\int _0^{2 \pi }x^2 \text{Sin}(\theta ) e^{i q x \text{Cos}(\theta )} \times \nonumber \\
&& \psi ^{n'l'm'}(x,\theta ,\phi )^*\psi ^{nlm}(x,\theta ,\phi )d\phi d\theta dx \nonumber,
\eea
where $\psi$'s are the hydrogen-like atomic wave-functions for TM, $q$ is the magnitude of the 3-vector part of the momentum transfer and $x=\frac{r}{a}$
with $a$ the Bohr radius for TM. $nl$ labels the incoming TM state (i.e. $1^3S_1$, $2^3S_1$...etc).
We now have the electric differential cross-sections for transition into other bound states or break-up given by,
\bea
\text{d$\sigma $}^{nl}_{\text{tot}}&=& Z^2 \frac{\alpha ^2 }{\pi }\left(1-F^{nl0,n'l'0}(q)\right) \nonumber \\
&\times& \frac{1}{a^2}|\Delta (q,Z)|^2q \text{d}q. \label{eq:tot}
\eea
Here $\Delta$ is the photon propagator in Thomas-Fermi-Molier form:
\begin{equation}
\Delta (q,Z) = 4 \pi \sum^3_{i=1} \frac{\alpha_i}{q^2 + \beta_i^2},
\end{equation}
where $\beta_i = \frac{m_e b_i}{121}Z^{1/3}$, with 
\begin{align}
	b_1 = 6.0, b_2 = 1.2, b_3 = 0.3,\nonumber \\
	\alpha_1 = 0.10, \alpha_2 = 0.55, \alpha_3 = 0.35 \nonumber
\end{align}
The electric differential cross-sections for transition from the $nl$ state to a specific $n'l'$ state is, 
\bea
\text{d$\sigma $}^{nl, n'l'} &=& \left(1-(-1)^{l-l'}\right)Z^2 \frac{\alpha ^2 }{\pi } \frac{1}{a^2}q |\Delta (q,Z) \nonumber \\
&\times& F^{nl0,n'l'0}\left(\frac{q}{2}\right)|^2\text{d}q
\label{eq:excitation}
\eea

To get total cross-sections, we integrate (\ref{eq:tot}) and (\ref{eq:excitation}) for $q\in\left[0,\infty \right)$.
We can then calculate the dissociation cross-section by subtracting from the total ``excitation cross-section''
(integrated (\ref{eq:tot})) the cross-sections for excitations into bound states (integrated (\ref{eq:excitation})). 
Selected results are summarized in Table 1.  
These results agree with those listed in \cite{Denisenko:1987gr}.
All results are calculated using the same formalism. 
The estimation of uncertainty for these excitation reactions is not straightforward, but a rough estimate can be derived by comparing to 
positronium results computed in the same formalism versus other methods.  
For example, the difference between results from \cite{Nemenov:1989nx} and ours is below $10\%$.
\begin{table}[b]
\label{NullData}
\begin{center}
\begin{tabular}{|c||c|c|c|}
\hline
& from 1S & from 2S & from 2P\\ 
\hline \hline
  to 1S & 0 & 0 & $3.20\times 10^{-20}$\\ 
 \hline
 to 2S & 0 & 0 & $3.82\times10^{-19}$\\ 
 \hline
 to 2P & $3.20\times 10^{-20}$ & $3.82\times10^{-19}$ & 0 \\ 
 \hline
  to 3S & 0 & 0 & $9.53\times10^{-21}$\\ 
 \hline
  to $e^{\!+}\!\!e^{\!-}$ & $2.65\times10^{-20}$ & $7.52\times10^{-20}$ & $3.34\times10^{-19}$\\ 
 \hline
  to X & $6.89\times10^{-20}$ & $5.92\times10^{-19}$ & $7.61\times10^{-19}$\\ 
 \hline
\end{tabular}
\caption{Cross-sections for excitations and dissociation (in $\text{cm}^2$) for Lead.
``to $e^{\!+}\!\!e^{\!-}$'' refers to dissociation of TM, while ``X'' refers to all final states inclusive.}
\end{center}
\end{table}

\section{Discovery Prospects and Conclusion}\label{sec:conclusion}

We're now ready to assess the feasibility of TM discovery and possible measurements
using upcoming fixed-target experiments. 
Typical experimental configurations consist
of beam currents of a few hundred nA of continuous-wave electrons (in the case of HPS)
with run times of O(months) and targets of thickness $\lesssim 1 \%$ radiation lengths. 
Using our new calculation of total TM relative yields, we can express total TM
production in terms of the number of electrons on target, assuming
a target thicker than a few dissociation lengths of $1^3S_1$, which is typically
$\lesssim 10^{-3}$ radiation lengths. 
For reference values of $I=450$ nA and 1 month of beam time,
the total number $N(x_c,l_c)$ of TM produced with $x>x_c$ and $l>l_c$ is,
\be
N(x_c,l_c) = Y(x_c,l_c) (7.27\times10^{18}) \left( \frac{I\times T}{450 \ nA\times month} \right)
\ee
For Lead (Tungsten is similar), we obtain total yields in the range of $200$ to $600$
events, depending on $x_c$ and $l_c$. With total acceptance and reconstruction 
efficiencies in the range of $10-20 \%$, this gives sizable detected yields. 

In this work, we have established that initial discovery of TM with the above experimental 
parameters should be possible.
Moreover, TM yields should be large enough that measuring basic TM properties may 
be possible.
In an upcoming paper~\cite{Upcoming}, a detailed analysis of target material,
target configuration, and detector layout will be presented in order to define an optimal 
strategy for measuring TM production cross-sections, lifetimes, and dissociation rates. 
Additional work related to TM can be found in \cite{Jentschura:1997tv,Jentschura:1997ma,Hughes:1960zz,Kuraev:1977rp,Karshenboim:1998am,Ginzburg:1998df,Nemenov:1972ph}. 

\noindent
{ \bf Acknowledgements:}
We thank Rouven Essig, Sarah Phillips, Maxim Pospelov, Natalia Toro, and Itay Yavin for 
numerous illuminating discussions and helpful feedback. 
We also thank Stanislaw Mrowczynski for guidance regarding dissociation physics. 


\end{document}